\documentclass[manuscript]{aastex}

\usepackage{rotating}

\slugcomment{Based on observations collected at the Observatoire de Haute-Provence (France).}

\shorttitle{Spectroscopic study of Of$^+$ supergiants}
\shortauthors{De Becker et al.}

\begin{document}

\title{Spectroscopic study of the variability of three northern Of$^+$ supergiants\slugcomment{Based on observations collected at the Observatoire de Haute-Provence (France).}}

\shorttitle{Spectroscopic study of Of$^+$ supergiants}
\shortauthors{De Becker et al.}

\author{M. De Becker\altaffilmark{1}, G. Rauw and N. Linder}
\affil{Institut d'Astrophysique et de G\'eophysique, Universit\'e de Li\`ege, 17, All\'ee du 6 Ao\^ut, B5c, B-4000 Sart Tilman, Belgium}
\email{debecker@astro.ulg.ac.be}

\altaffiltext{1}{Postdosctoral Researcher FRS/FNRS (Belgium)}

\begin{abstract}
The transition from early Of stars to WN type objects is poorly understood. O-type supergiants with emission lines (OIf$^+$) are considered to be intermediate between these two classes. The scope of this paper is to investigate the spectral variability of three Of$^+$ supergiants. We constituted spectral time series of unprecedented quality for our targets ($\sim$\,200 spectra in total), essentially in the blue domain, covering time-scales from a few hours up to a few years. Time Variance Spectrum (TVS) and Fourier analyses were performed in order to characterize their spectral variability. We report on a correlated significant line profile variability in the prominent He\,{\sc ii} $\lambda$ 4686 and H\,$\beta$ lines most likely related to the strong stellar winds. The variability pattern is similar for the three stars investigated (HD\,14947, HD\,15570 and HD\,16691), and the main differences are more quantitative than qualitative. However, the reported time-scales are somewhat different, and the most striking variability pattern is reported for HD\,16691. We did not find any clear evidence for binarity, and we focus mainly on an interpretation based on a single star scenario. We show that the behaviour of the three stars investigated in this study present strong similarities, pointing to a putative common scenario, even though a few differences should be noted. Our preferred interpretation scheme is that of Large Scale Corotating Structures modulating the profile of the lines that are produced in the strong stellar wind.
\end{abstract}


\maketitle

\section{Introduction} \label{intro}

In the standard evolutionary scheme, Of stars are the progenitors of Wolf-Rayet stars. As the star evolves into the supergiant phase, its stellar wind becomes stronger, which translates in some lines appearing in strong emission in the optical spectrum (He\,{\sc ii} $\lambda$ 4686, H\,$\alpha$...). On the other hand, WR stars are characterized by strong emission lines, also related to their strong and dense winds. The similarities in the spectral morphology in the optical domain of these two classes of stars is a well-established fact for several decades \citep{contiliege}. In this context, \citet{contitrans} discussed the similarities between the K-band spectra of the O4If$^+$ stars HD\,16691 and HD\,190429A, and those of Wolf-Rayet stars. The latter authors pointed out the {\it connection in the spectral morphology between some early Of and WN stars} likely attributable to an evolutionary relationship. Another star worth considering in the same framework is the bright O4If$^+$ star HD\,15570, member of the open cluster IC\,1805, that has been qualified as transitional by \citet{ws15570} on the basis of its UV spectral properties. In addition, \citet{cappabubble} reported on the probable presence around HD\,16691 and HD\,14947 (O5If$^+$) of interstellar bubbles that are very similar in shape to other bubbles surrounding WR stars.

\begin{figure*}
\centering
\includegraphics[width=170mm]{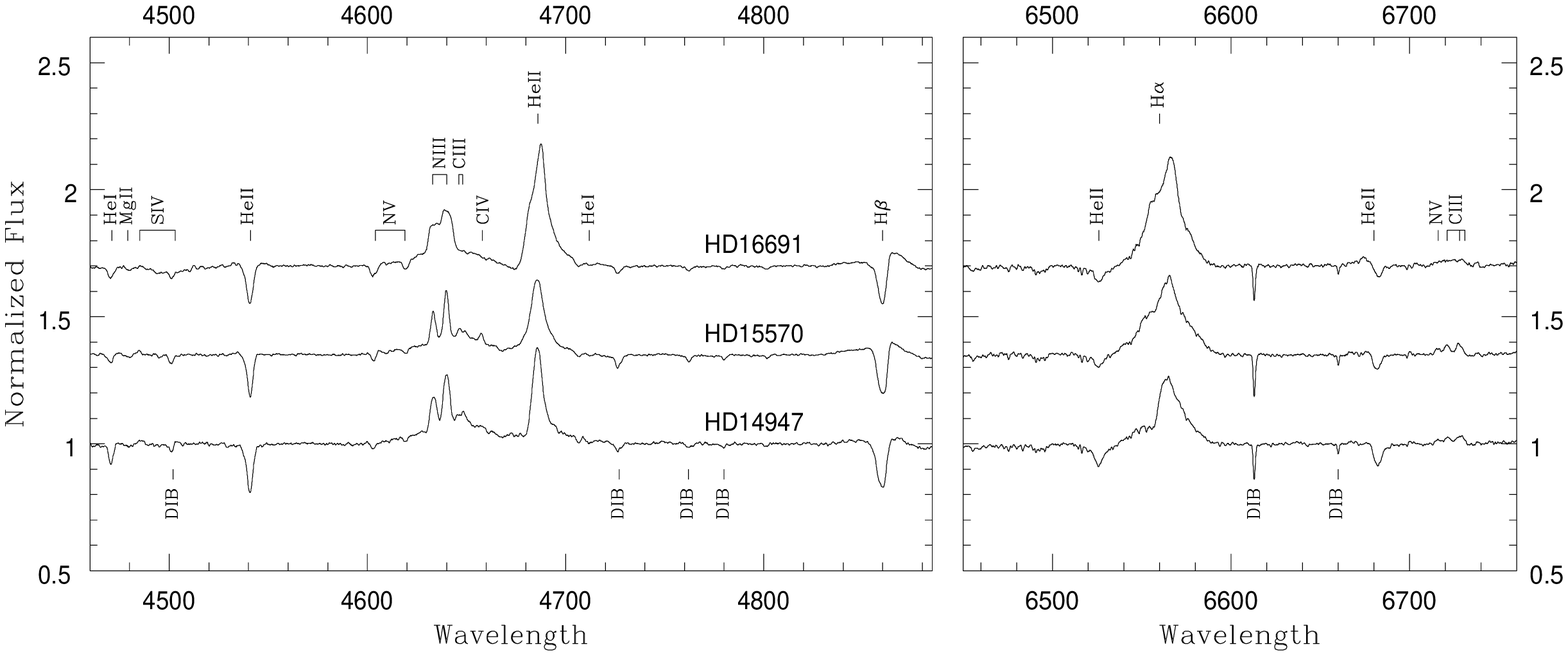}
\caption{Mean visible spectra of HD\,14947, HD\,15570 and HD\,16691 between 4460 and 4890\,\AA\, (left part) and between 6450 and 6760 \AA\, (right part).}
\label{specsg}
\end{figure*}

If one wants to address the issue of the dynamics of the quite extreme stellar winds of such transitional objects, variability studies are likely to bring crucial information. In their study of the line profile variations of H\,$\alpha$, \citet{markova-lpv} reported on significant variations in the case of HD\,190429A, HD\,14947 and HD\,16691. Even though their time series were limited to only a few spectra, they pointed out some similarities in the spectral shape and in the behaviour of these stars. They also noted that the lack of significant line profile variability in other lines such as He\,{\sc ii} $\lambda$ 6527\,\AA\, suggests that the variations in H\,$\alpha$ were probably due to processes in the wind. However, to investigate in detail such a line profile variability, a very good time sampling is required.

In the context of the present study, we focus on three targets: HD\,14947, HD\,16691 and HD\,15570. The latter one has already been investigated by \citet{jenam1805}, but the time series was not well suited for the study of short time scale variations, i.e. of the order of the day or shorter. The significantly improved time series presented here are used to more adequately characterize the line profile variability of our three targets. We first discuss the optical spectrum of the stars (Section\,\ref{spec}). The main part of the paper consists in a detailed description of the variability study (Section\,\ref{res}), followed by a discussion in Section\,\ref{disc}. Our conclusions are presented in Section\,\ref{conc}. 

\section{Observations and data reduction}\label{obs}
Spectroscopic observations were collected at the Observatoire de Haute-Provence (OHP, France) during several observing runs from October 2004 to Autumn 2007. All spectra were obtained with the Aur\'elie spectrograph fed by the 1.52\,m telescope \citep{aurelie}, using the same setup as described for instance by \citet{Let8a,ic1805_2}. Our collection of spectra is described in Table\,\ref{runs}, and the detailed journal of the observations is given in Table\,\ref{log}. The typical exposure time was about 30 -- 45 minutes, depending on the weather conditions. In order to carry out a detailed variability study of our targets, we organized the observing campaign in such a way that we covered time scales from a few hours up to a few years. In total, we obtained 57, 64 and 64 blue spectra between 4450 and 4900\,\AA\, respectively for HD\,14947, HD\,15570 and HD\,16691. In addition, in October 2005, we otained 7, 5 and 6 red spectra between 6340 and 6780\,\AA\, respectively for the three targets spread over a few days. The resolving power is about 8000 in the blue, and 11000 in the red. All data were treated following the reduction procedure already described by \citet{ic1805_1}. 

\begin{table}
\caption{Description of observing runs for the blue domain. The first and second columns give the name of the campaign as used in the text as well as the instrumentation used. For each star, the next columns yield the number of spectra obtained, the time elapsed between the first and the last spectrum of the run ($\Delta$T), the natural width ($\Delta\nu_\mathrm{nat}$) of a peak of the power spectrum taken as 1/$\Delta$T, and finally the mean signal-to-noise ratio of each data set. In the case of HD\,15570, the information concerning observations before 2007 can be found in \citet{ic1805_2}.\label{runs}}
\begin{center}
\begin{tabular}{l c c c c}
\hline\hline
 Obs. run &  N & $\Delta$T & $\Delta\nu_\mathrm{nat}$ & S/N \\
  &   & (d) & (d$^{-1}$) & \\
\hline
\multicolumn{2}{l}{HD\,14947} & & & \\
\hline
\vspace*{-0.2cm}\\
Oct.2004 & 4 & 10.1 & 0.99 & 360 \\
Oct.2005 & 4 & 6.0 & 0.17 & 320 \\
Sept.2006 & 2 & 2.0 & 0.50 & 290 \\
Oct.2006 & 5 & 1.2 & 0.85 & 270 \\
Aut.2007 & 42 & 26.1 & 0.04 & 300 \\
\vspace*{-0.2cm}\\
\hline
\multicolumn{2}{l}{HD\,15570} & & & \\
\hline
\vspace*{-0.2cm}\\
Aut.2007 & 28 & 11.0 & 0.09 & 300 \\
\vspace*{-0.2cm}\\
\hline
\multicolumn{2}{l}{HD\,16691} & & & \\
\hline
\vspace*{-0.2cm}\\
Oct.2004 & 5 & 10.0 & 0.10 & 300 \\
Oct.2005 & 4 & 6.0 & 0.17 & 270 \\
Sept.2006 & 2 & 4.0 & 0.25 & 210 \\
Oct.2006 & 7 & 1.2 & 0.85 & 200 \\
Aut.2007 & 46 & 26.0 & 0.04 & 300 \\
\vspace*{-0.2cm}\\
\hline
\end{tabular}
\end{center}
\end{table}

During this campaign, particular efforts were devoted to the sampling of many time-scales. As this is a critical issue in variability studies, we managed to collect as much data as possible with various time intervals between two consecutive observations. Table\,\ref{log} shows that the campaign spreads over more than three years for HD\,14947 and HD\,16691, and more than 7 years in the case of HD\,15570, with several observing runs in between. Moreover, our targets were oberved several times a night, over several nights, in order to cover shorter time-scales. We note that this observing campaign ended in October-November 2007, with a 28 consecutive nights of observation. The latter observing run allowed us to obtain a very good sampling of shorter time-scales, that turned out to be critical in the context of the variability study of our targets. As we were intending to study the variability in line profiles, we focused on high quality spectra with high signal-to-noise ratios, in most cases significantly higher than 200 (see Table\,\ref{runs}).

\begin{sidewaystable}
\caption{Journal of the observations for HD\,14947, HD\,15570 and HD\,16691. The heliocentric Julian date is given as HJD - 2\,450\,000, and the date (yyyy/mm/dd) is that of the beginning of the night. In each case, the spectral domain (blue or red) is specified (available in electronic form only). \label{log}}
\begin{center}
\begin{tabular}{l | c c c | c c c | c c c}
\hline\hline
  & \multicolumn{3}{c}{HD\,14947} & \multicolumn{3}{c}{HD\,15570} & \multicolumn{3}{c}{HD\,16691} \\
\vspace*{-0.2cm}\\
\hline
  & HJD  & Date & Domain & HJD  & Date & Domain & HJD  & Date & Domain \\
\vspace*{-0.2cm}\\
\hline
1  & 3286.560 & 2004/10/07 & blue & 1811.640  & 2000/10/23 & blue & 3286.586 & 2004/10/07 & blue \\
2  & 3289.681 & 2004/10/10 & blue & 1812.638  & 2000/10/24 & blue & 3289.647 & 2004/10/10 & blue \\
3  & 3290.624 & 2004/10/11 & blue & 1813.670  & 2000/10/25 & blue & 3290.581 & 2004/10/11 & blue \\
4  & 3296.666 & 2004/10/17 & blue & 1814.665  & 2000/10/26 & blue & 3295.616 & 2004/10/16 & blue \\
5  & 3648.483 & 2005/10/04 & red & 1815.666  & 2000/10/27 & blue & 3296.624 & 2004/10/17 & blue \\
6  & 3648.652 & 2005/10/04 & blue & 1819.606  & 2000/10/01 & blue & 3648.529 & 2005/10/04 & red \\
7  & 3649.454 & 2005/10/05 & red & 1821.631  & 2000/10/03 & blue & 3648.588 & 2005/10/04 & blue \\
8  & 3650.385 & 2005/10/06 & red & 2163.604  & 2001/09/10 & blue & 3649.641 & 2005/10/05 & red \\
9  & 3650.619 & 2005/10/06 & red & 2164.635  & 2001/09/11 & blue & 3650.567 & 2005/10/06 & red \\
10  & 3652.458 & 2005/10/08 & red & 2165.609  & 2001/09/12 & blue & 3652.374 & 2005/10/08 & red \\
11  & 3652.614 & 2005/10/08 & blue & 2167.583  & 2001/09/14 & blue & 3652.500 & 2005/10/08 & red \\
12  & 3653.366 & 2005/10/09 & red & 2170.626  & 2001/09/17 & blue & 3652.555 & 2005/10/08 & blue \\
13  & 3654.378 & 2005/10/10 & red & 2518.603  & 2002/08/31 & blue & 3653.407 & 2005/10/09 & red \\
14  & 3654.433 & 2005/10/10 & blue & 2520.545  & 2002/09/02 & blue & 3653.460 & 2005/10/09 & blue \\
15  & 3654.605 & 2005/10/10 & blue & 2523.630  & 2002/09/05 & blue & 3654.635 & 2005/10/10 & blue \\
\vspace*{-0.2cm}\\
\hline
\end{tabular}
\end{center}
\end{sidewaystable}

\setcounter{table}{1}

\begin{sidewaystable}
\caption{(continued)}
\begin{center}
\begin{tabular}{l | c c c | c c c | c c c}
\hline\hline
16  & 3982.643 & 2006/09/03 & blue & 2524.585  & 2002/09/06 & blue & 3980.610 & 2006/09/01 & blue \\
17  & 3984.629 & 2006/09/05 & blue & 2527.594  & 2002/09/09 & blue & 3984.602 & 2006/09/05 & blue \\
18  & 4033.468 & 2006/10/24 & blue & 2528.602  & 2002/09/10 & blue & 4033.436 & 2006/10/24 & blue \\
19  & 4033.553 & 2006/10/24 & blue & 2529.584  & 2002/09/11 & blue & 4033.524 & 2006/10/24 & blue \\
20  & 4033.664 & 2006/10/24 & blue & 2531.594  & 2002/09/13 & blue & 4033.631 & 2006/10/24 & blue \\
21  & 4034.383 & 2006/10/25 & blue & 2532.593  & 2002/09/14 & blue & 4034.437 & 2006/10/25 & blue\\
22  & 4034.643 & 2006/10/25 & blue & 2533.546  & 2002/09/15 & blue & 4034.494 & 2006/10/25 & blue \\
23  & 4397.373 & 2007/10/23 & blue & 2916.650  & 2003/10/03 & blue & 4034.555 & 2006/10/25 & blue \\
24  & 4397.432 & 2007/10/23 & blue & 2918.595  & 2003/10/05 & blue & 4034.614 & 2006/10/25 & blue \\
25  & 4397.491 & 2007/10/23 & blue & 2919.577  & 2003/10/06 & blue & 4397.343 & 2007/10/23 & blue \\
26  & 4397.568 & 2007/10/23 & blue & 2922.609  & 2003/10/09 & blue & 4397.404 & 2007/10/23 & blue \\
27  & 4401.342 & 2007/10/27 & blue & 2925.620  & 2003/10/12 & blue & 4397.461 & 2007/10/23 & blue \\
28  & 4401.401 & 2007/10/27 & blue & 2928.616  & 2003/10/15 & blue & 4397.542 & 2007/10/23 & blue \\
29  & 4401.461 & 2007/10/27 & blue & 2934.620  & 2003/10/21 & blue & 4397.596 & 2007/10/23 & blue \\
30  & 4401.543 & 2007/10/27 & blue & 3286.612  & 2004/10/07 & blue & 4400.412 & 2007/10/26 & blue \\
31  & 4402.360 & 2007/10/28 & blue & 3287.487  & 2004/10/08 & blue & 4400.484 & 2007/10/26 & blue \\
32  & 4402.415 & 2007/10/28 & blue & 3289.518  & 2004/10/10 & blue & 4400.542 & 2007/10/26 & blue \\
33  & 4402.470 & 2007/10/28 & blue & 3290.510  & 2004/10/11 & blue & 4400.595 & 2007/10/26 & blue \\
34  & 4402.561 & 2007/10/28 & blue & 3294.506  & 2004/10/15 & blue & 4402.335 & 2007/10/28 & blue \\
35  & 4404.489 & 2007/10/30 & blue & 3295.519  & 2004/10/16 & blue & 4402.348 & 2007/10/28 & blue \\
36  & 4405.371 & 2007/10/31 & blue & 3296.509  & 2004/10/17 & blue & 4402.442 & 2007/10/28 & blue \\
37  & 4405.421 & 2007/10/31 & blue & 3648.506  & 2005/10/04 & red & 4402.499 & 2007/10/28 & blue \\
\hline
\end{tabular}
\end{center}
\end{sidewaystable}

\setcounter{table}{1}

\begin{sidewaystable}
\caption{(continued)}
\begin{center}
\begin{tabular}{l | c c c | c c c | c c c}
\hline\hline
38  & 4405.470 & 2007/10/31 & blue & 3649.612  & 2005/10/05 & red & 4402.534 & 2007/10/28 & blue \\
39  & 4405.547 & 2007/10/31 & blue & 3650.407  & 2005/10/06 & red & 4403.391 & 2007/10/29 & blue \\
40  & 4408.364 & 2007/11/03 & blue & 3652.478  & 2005/10/08 & red & 4403.443 & 2007/10/29 & blue \\
41  & 4408.419 & 2007/11/03 & blue & 3653.386  & 2005/10/09 & red & 4403.494 & 2007/10/29 & blue \\
42  & 4408.440 & 2007/11/03 & blue & 4400.455  & 2007/10/26 & blue & 4403.580 & 2007/10/29 & blue \\
43  & 4408.491 & 2007/11/03 & blue & 4400.513  & 2007/10/26 & blue & 4405.346 & 2007/10/31 & blue \\
44  & 4408.569 & 2007/11/03 & blue & 4400.570  & 2007/10/26 & blue & 4405.395 & 2007/10/31 & blue \\
45  & 4409.391 & 2007/11/04 & blue & 4400.622  & 2007/10/26 & blue & 4405.445 & 2007/10/31 & blue \\
46  & 4409.445 & 2007/11/04 & blue & 4401.370  & 2007/10/27 & blue & 4405.497 & 2007/10/31 & blue \\
47  & 4409.550 & 2007/11/04 & blue & 4401.430  & 2007/10/27 & blue & 4405.522 & 2007/10/31 & blue \\
48  & 4409.571 & 2007/11/04 & blue & 4401.489  & 2007/10/27 & blue & 4405.572 & 2007/10/31 & blue \\
49  & 4411.380 & 2007/11/06 & blue & 4401.571  & 2007/10/27 & blue & 4406.336 & 2007/11/01 & blue \\
50  & 4411.449 & 2007/11/06 & blue & 4403.418  & 2007/10/29 & blue & 4406.452 & 2007/11/01 & blue \\
51  & 4411.586 & 2007/11/06 & blue & 4403.468  & 2007/10/29 & blue & 4406.504 & 2007/11/01 & blue \\
52  & 4412.405 & 2007/11/07 & blue & 4403.553  & 2007/10/29 & blue & 4406.563 & 2007/11/01 & blue \\
53  & 4413.405 & 2007/11/08 & blue & 4404.401  & 2007/10/30 & blue & 4409.363 & 2007/11/04 & blue \\
54  & 4413.432 & 2007/11/08 & blue & 4404.458  & 2007/10/30 & blue & 4409.418 & 2007/11/04 & blue \\
55  & 4413.547 & 2007/11/08 & blue & 4404.519  & 2007/10/30 & blue & 4409.472 & 2007/11/04 & blue \\
56  & 4414.484 & 2007/11/09 & blue & 4406.363  & 2007/11/01 & blue & 4409.502 & 2007/11/04 & blue \\
57  & 4414.575 & 2007/11/09 & blue & 4406.425  & 2007/11/01 & blue & 4409.599 & 2007/11/04 & blue \\
58  & 4416.374 & 2007/11/11 & blue & 4406.478  & 2007/11/01 & blue & 4410.393 & 2007/11/05 & blue \\
59  & 4417.554 & 2007/11/12 & blue & 4406.532  & 2007/11/01 & blue & 4410.449 & 2007/11/05 & blue \\
\vspace*{-0.2cm}\\
\hline
\end{tabular}
\end{center}
\end{sidewaystable}

\setcounter{table}{1}

\begin{sidewaystable}
\caption{(continued)}
\begin{center}
\begin{tabular}{l | c c c | c c c | c c c}
\hline\hline
60  & 4417.579 & 2007/11/12 & blue & 4408.392  & 2007/11/03 & blue & 4410.500 & 2007/11/05 & blue \\
61  & 4419.442 & 2007/11/14 & blue & 4408.464  & 2007/11/03 & blue & 4410.611 & 2007/11/05 & blue \\
62  & 4419.471 & 2007/11/14 & blue & 4408.517  & 2007/11/03 & blue & 4413.375 & 2007/11/08 & blue \\
63  & 4423.374 & 2007/11/18 & blue & 4408.542  & 2007/11/03 & blue & 4413.509 & 2007/11/08 & blue \\
64  & 4423.503 & 2007/11/18 & blue & 4410.362  & 2007/11/05 & blue & 4416.527 & 2007/11/11 & blue \\
65  &   &  &  & 4410.420  & 2007/11/05 & blue & 4416.590 & 2007/11/11 & blue \\
66  &   &  &  & 4410.528  & 2007/11/05 & blue & 4417.437 & 2007/11/12 & blue \\
67  &   &  &  & 4410.553  & 2007/11/05 & blue & 4417.497 & 2007/11/12 & blue \\
68  &   &  &  & 4411.414  & 2007/11/06 & blue & 4417.525 & 2007/11/12 & blue \\
69  &   &  &  & 4411.500  & 2007/11/06 & blue & 4419.519 & 2007/11/14 & blue \\
70  &   &  &  &  &  &  & 4423.349 & 2007/11/18 & blue \\
\vspace*{-0.2cm}\\
\hline
\end{tabular}
\end{center}
\end{sidewaystable}

\section{The optical spectrum\label{spec}} 
We plot the mean spectra of our three targets in Figure\,\ref{specsg}. These spectra are in excellent agreement with the spectral types and luminosity classes reported in the litterature (see Section\,\ref{intro}). The main absorption and emission lines are labelled in the figure. We note that, in the case of HD\,15570, a spectrum covering the whole visible domain is given in \citet{ic1805_2}.

When the spectra of our three targets are compared, several common features are worth mentioning:
\begin{enumerate}
\item[-] the blue spectrum is dominated by the strong, broad, and asymmetric emission of He\,{\sc ii} $\lambda$ 4686,
\item[-] H\,$\alpha$ is in strong emission, with a shape similar to that of He\,{\sc ii} $\lambda$ 4686,
\item[-] the second strongest emission lines in the blue spectrum are N\,{\sc iii} $\lambda\lambda$ 4634,4641,
\item[-] the spectral region between about 4600 and 4720\,\AA\, presents a broad emission bump,
\item[-] H\,$\beta$ displays a clear P\,Cygni shape, with an additionnal very weak emission shoulder on the blue side of the absorption component\footnote{This weak emission on the blue side of H\,$\beta$ is most probably a residual from the broad emission component of the P\,Cygni profile that is not completely compensated by the narrower blue shifted absorption. A broad emission component from the wind is indeed expected, considering the width of other lines originating from the stellar wind such as He\,{\sc ii} $\lambda$ 4686 and H\,$\alpha$.}.
\end{enumerate}

Beside these similarities, it should be pointed out that the strong emission lines (He\,{\sc ii} and H\,$\alpha$) are significantly stronger in the case of HD\,16691, as compared to the two other stars. The N\,{\sc iii} emission doublet of HD\,16691 presents also a rather confused shape with respect to the well individualized ones observed in HD\,14947 and HD\,15570. In addition, a slightly more complex profile is observed for H\,$\beta$ for the same star, with a weak additional emission component on the top of the P\,Cygni profile. The latter feature is not obvious in the mean spectrum shown in Figure\,\ref{specsg}, but is observed in some profiles presented in Figure\,\ref{tvs}. The C\,{\sc iii} $\lambda\lambda$ 4647,4650 and C\,{\sc iv} $\lambda$ 4660 features are much less pronounced in the case of HD\,16691, but the N\,{\sc v} $\lambda\lambda$ 4606,4619 absorption lines are somewhat stronger than in the case of HD\,14947 and HD\,15570.

\section{Results \label{res}}

\begin{figure*}
\centering
\includegraphics[width=140mm]{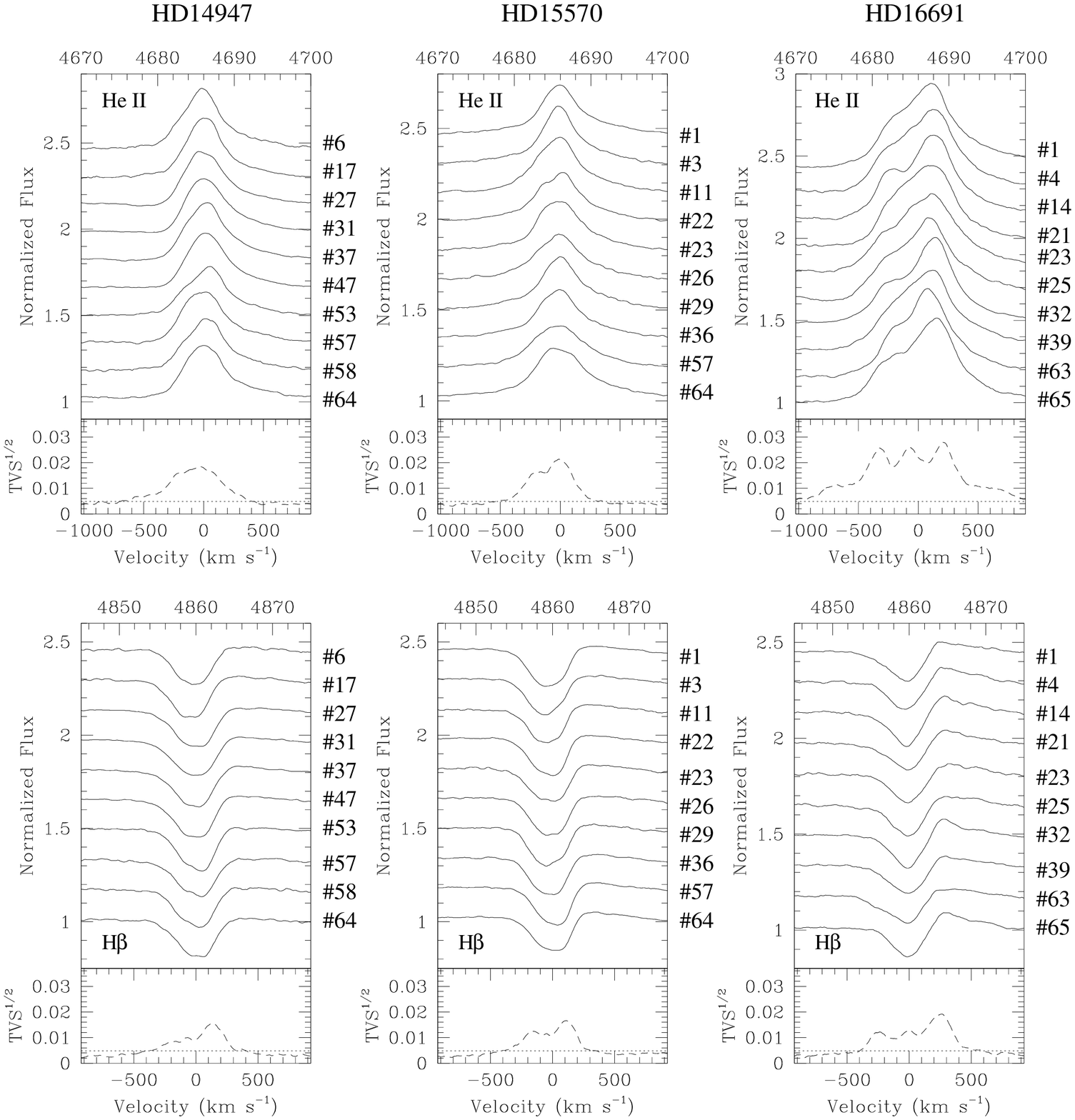}
\caption{Profile variability of the strongest lines respectively for HD\,14947, HD\,15570 and HD\,16691 (from the left to the right). {\it Upper panels:} He\,{\sc ii} $\lambda$ 4686. {\it Lower panels:} H\,$\beta$. In each case, the upper plot shows a sample of profiles, and the lower one provides the TVS for the complete time series. The numbers on the left of the individual profiles correspond to those attributed in Table\,\ref{log}. The horizontal dotted line in the lower panels stands for the 99\,$\%$ confidence level. In order to give an idea of the relative variability level of every line of each star, the TVS scale is the same in all panels.}
\label{tvs}
\end{figure*}

\begin{figure*}
\centering
\includegraphics[width=140mm]{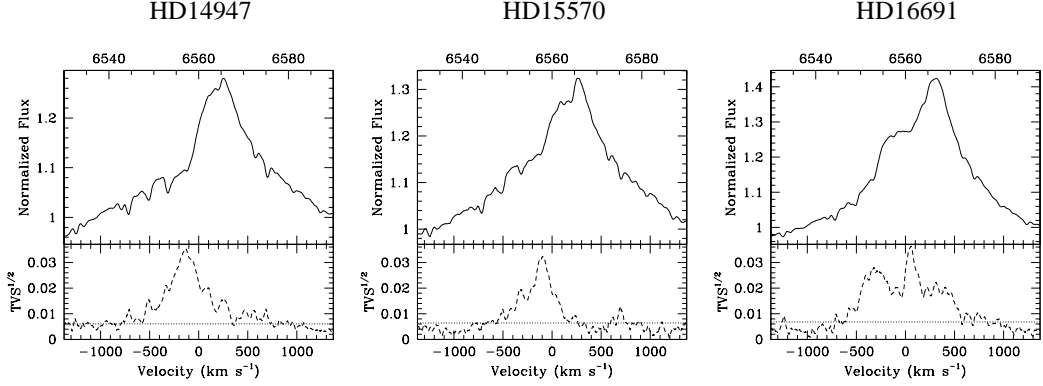}
\caption{Profile variability of the H\,$\alpha$ line respectively for HD\,14947, HD\,15570 and HD\,16691 (from the left to the right). In each case, the upper plot shows the mean profile, and the lower one provides the TVS for the October 2005 time series in the red domain.}
\label{tvsred}
\end{figure*}

We first searched for significant line profile variations using the Temporal Variance Spectrum (TVS) technique as described by \citet{fullertontvs}, and previously applied for instance by \citet{oef2}. This technique was first applied to the complete time series. For the three stars, very significant variations are detected mostly in the case of the He\,{\sc ii} $\lambda$ 4686, H\,$\beta$ and H\,$\alpha$ lines (at the 99\,$\%$ confidence level). The profiles of these lines, along with the square root of the TVS, are represented for HD\,14947, HD\,15570 and HD\,16691 in Figures \ref{tvs} and \ref{tvsred}. For all stars, the variability level of the He\,{\sc ii} and H\,$\alpha$ lines is higher than that of H\,$\beta$ (see below for more detailed individual discussions). We insist on the fact that the TVS of H\,$\alpha$ is computed from a limited number of spectra all collected in October 2005 (see Table\,\ref{log}). In addition, we note that the behaviour of He\,{\sc ii} $\lambda$ 4686 and H\,$\beta$ presents strong similarities, as illustrated by the variations of the asymmetric profiles in Figure\,\ref{tvs}. 

We used the generalized Fourier technique described by \citet{HMM}, and revised by \citet{gosset30a}. This technique is especially adapted to the case of unequally spaced data. We systematically applied this technique to our spectral time series, in order to search for possible frequencies (or more specifically the corresponding time-scales) ruling the detected variations. Our approach consisted in computing the power spectrum at each wavelength step leading to a two-dimensioned power spectrum. Mean periodograms were then computed to obtain an overview of the power as a function of frequency for the whole wavelength interval considered in the temporal analysis. We also applied prewhitening techniques considering candidate frequencies in order to check their capability to reproduce the spectral variations \citep[see e.g.][for applications of these techniques to line profile variability analyses]{oef2}. Most of the time, we separately considered the complete data set (about 60 spectra for each star in the blue domain), and the 2007 data set that offers an improved temporal sampling more adequate to search for short term variations (i.e. typically from several hours up to a few days). We note that the detailed temporal analyses were only performed on blue spectra. Our red spectra time series are indeed too limited and too sparse for this purpose. 

\subsection{HD\,14947 \label{res1}}

\begin{figure}
\centering
\includegraphics[width=80mm]{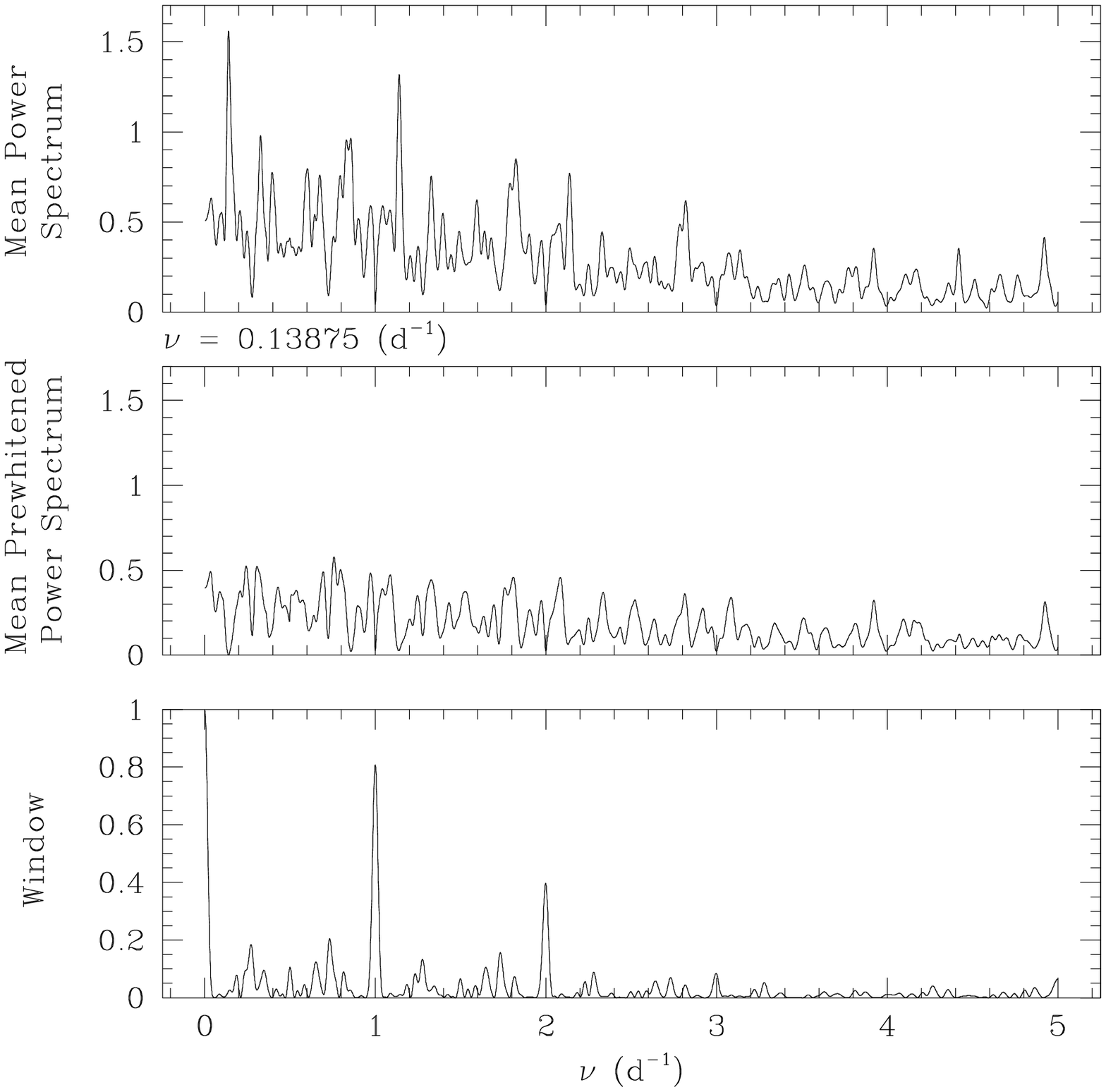}
\caption{Fourier analysis of the He\,{\sc ii} $\lambda$ 4686 line betwen 4678 and 4694\,\AA\, in the case of HD\,14947. {\it Upper panel:} mean power spectrum between 0 and 5\,d$^{-1}$. {\it Middle panel:} mean power spectrum obtained after prewhitening using the frequency of the highest peak of the initial power spectrum, i.e. 0.13875\,d$^{-1}$. {\it Lower panel:} spectral window directly related to the sampling of the time series.\label{fourhe14947}}
\end{figure}

\begin{figure}
\centering
\includegraphics[width=80mm]{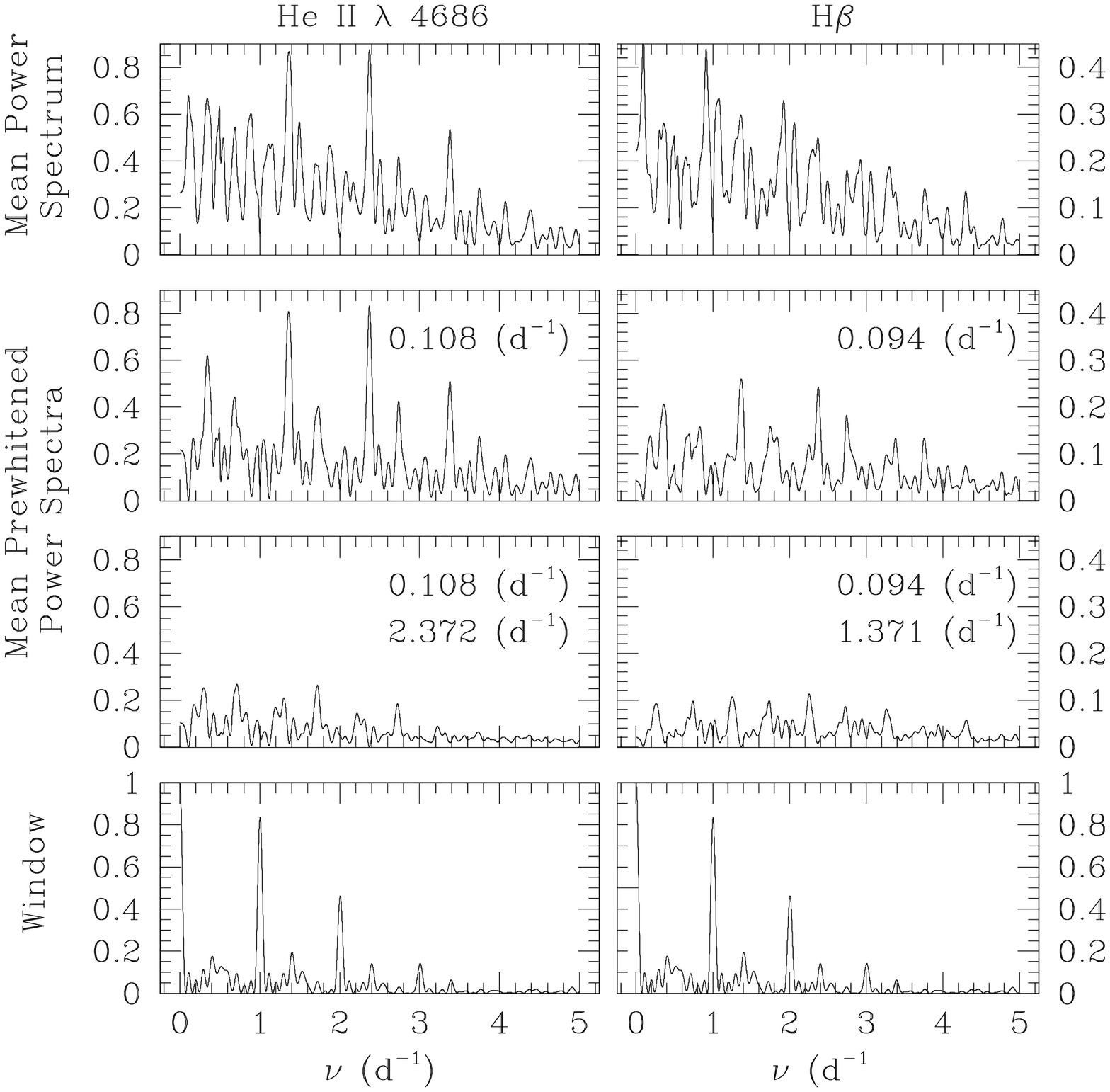}
\caption{Fourier analysis of the He\,{\sc ii} $\lambda$ 4686 line betwen 4678 and 4694\,\AA\, (left part) and of H\,$\beta$ between 4854 and 4866\,\AA\, (right part) in the case of HD\,15570. {\it Upper panel:} mean power spectra between 0 and 5\,d$^{-1}$. {\it Middle panels:} mean power spectrum obtained after prewhitening using the frequencies specified in each panel. {\it Lower panel:} spectral window directly related to the sampling of the time series, identical for both lines as it comes from the same times series.\label{fourheh15570}}
\end{figure}

Beside the variations in the profile of He\,{\sc ii} $\lambda$ 4686 and H\,$\beta$, the TVS analysis revealed also lower amplitude (though significant) variations in N\,{\sc iii} $\lambda\lambda$ 4634,4641 and in He\,{\sc ii} $\lambda$ 4542 (in absorption). We noted also marginal variations in He\,{\sc i} $\lambda$ 4471.

The Fourier technique applied to the complete data set (57 blue spectra) revealed a rather complex mean power spectrum, similar for all lines presenting significant variability. We report on highest peaks located at frequencies of about 0.14, 0.33 and 0.88\,d$^{-1}$ (the latter is likely at least partly an alias of 0.14\,d$^{-1}$), but it is quite difficult to discriminate between them in terms of dominating frequencies. We therefore focused mainly on the 2007 data set (42 blue spectra spread over $\sim$\,26 days), in order to get rid of potential long term trends (physical or not) likely to be present in long time series, and to isolate preferentially shorter time scales. The mean periodogram obtained for He\,{\sc ii} $\lambda$ 4686 on the 2007 time series is presented in Figure\,\ref{fourhe14947}. It was calculated between 4678 and 4694\,\AA\,, i.e. the wavelength interval where the variability is the stronger. As we obtained data with time intervals as short as one or two hours between two consecutive spectra, we were able to investigate frequencies as high as 10\,d$^{-1}$. However, as we did not detect any significant power at high frequencies, we limit the discussion to the frequency interval between 0 and 5\,d$^{-1}$. As can be seen in the upper panel of Figure\,\ref{fourhe14947}, the power spectrum is dominated by a frequency close to 0.139\,d$^{-1}$, corresponding to a time scale of the order of 7.2\,d. Prewhitening with the latter frequency leads to a rather good result, as shown in the middle panel of the same figure. However, some residual power is still present. The same analysis performed on the H\,$\beta$ line revealed a similar power spectrum, with the highest peak located at a frequency of 0.1425\,d$^{-1}$. In this case, much more residual power remains after prewhitening than in the case of the He\,{\sc ii} line.

\subsection{HD\,15570 \label{res2}}

A significant variability has already been reported in He\,{\sc ii} $\lambda$ 4686 \AA\,and H\,$\beta$ by \citet{jenam1805}. The latter analysis was based on a fraction of the data set discussed in this paper (up to 2004). The main improvement of the present study comes from the significantly better sampling of shorter time-scales thanks to the 28 spectra collected in 2007 (over about 11 days). We do not report on any significant variability for the other lines of the blue spectrum. However, very significant variations are detected for H\,$\alpha$, even though our temporal sampling of this line did not allow us to establish the time-scale of these variations. Significant variations of the H\,$\alpha$ line were already reported by \citet{polcaro}.

\begin{figure}
\centering
\includegraphics[width=80mm]{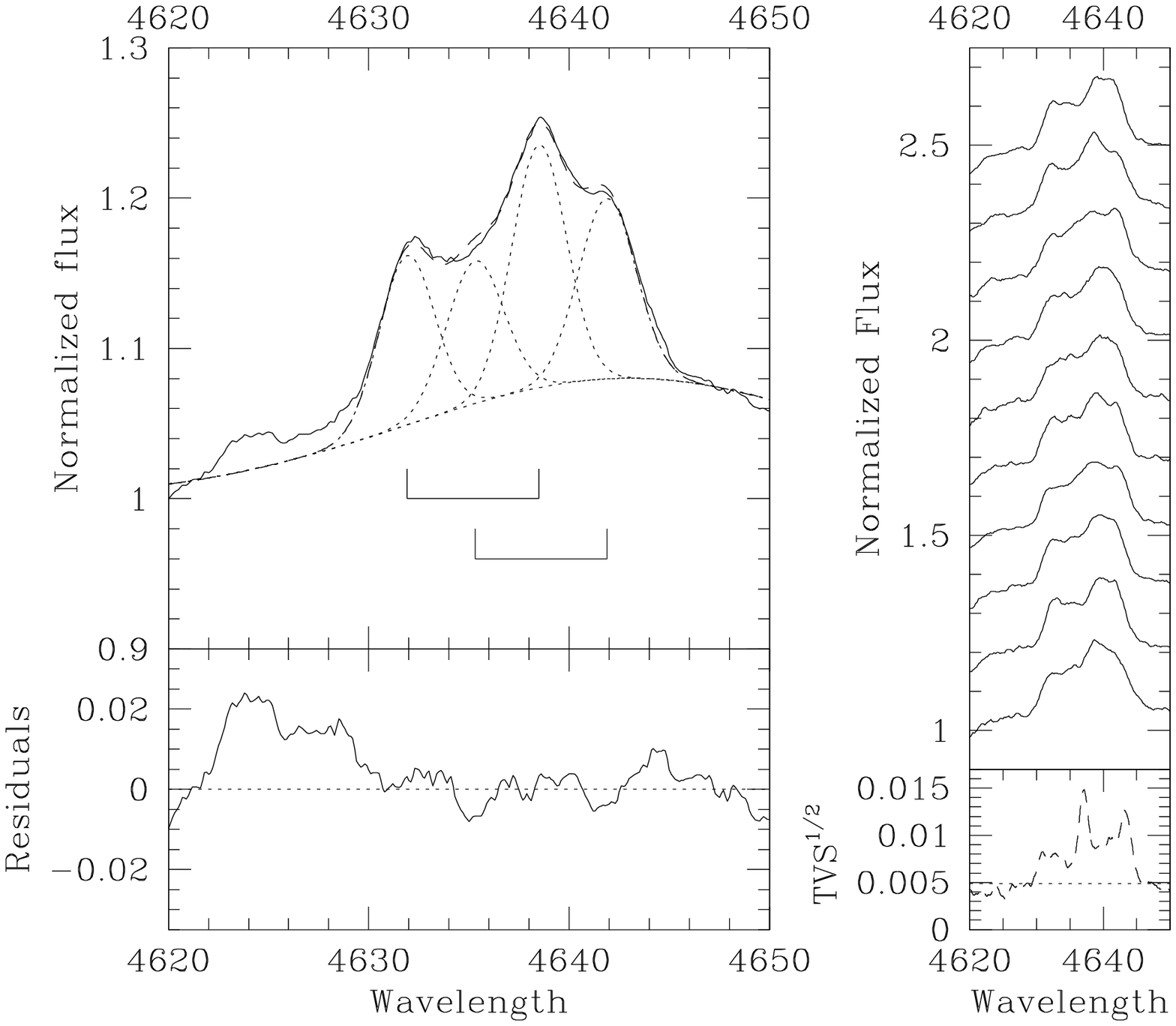}
\caption{N\,{\sc iii} line profile at 4634,4641\,\AA\, in the spectrum of HD\,16691. {\it Left part:} fit of the profile (spectrum \#4 in Table\,\ref{log}) using two pairs of Gaussians. The central positions of the four Gaussians correspond respectively to radial velocities of --136.5, 83.4, --147.3 and 72.3\,km\,s$^{-1}$. The lower panel gives the residual of the fit, in the sense data minus model. An additional broad component is used in order to model the emission bump already mentioned in Section\,\ref{spec}. {\it Right part:} variations in the N\,{\sc iii} line profile illustrated by a sample of spectra (the same as in Figure\,\ref{tvs}), and by the TVS of the complete time series in the lower panel.\label{nitrogen}}
\end{figure}

The power spectrum of the complete time series (64 blue spectra) is dominated by a peak located very close to a frequency of about 1\,d$^{-1}$ (and corresponding aliases), most probably related to the sampling frequency of the major part of the time series (mostly before 2007). A second family of peaks is also revealed, with a strongest peak located at a frequency close to 1.37\,d$^{-1}$. Once again, we clarified the situation by considering only the 2007 data set. In this case, the power spectrum is dominated by two components: a poorly constrained long term trend (with essentially its $1-\nu$ alias)\footnote{The frequency close to 0.1\,d$^{-1}$ may be related to the timespan of the data collected in 2007, of the order of 11 days.}, and a series of peaks at $\nu_i$ = $\nu_o$ + $i$ ($\nu_o$ being close to 0.37\,d$^{-1}$), with a highest peak for $i$ equal to 1 and to 2 respectively for H\,$\beta$ and He\,{\sc ii} $\lambda$ 4686 (along with their aliases at $1+i-\nu_i$). This situation is illustrated in Figure\,\ref{fourheh15570}, where the result of successive prewhitening with two frequencies is shown for He\,{\sc ii} $\lambda$ 4686 (left part) and H\,$\beta$ (right part). Even though we scrupulously selected the highest peaks to prepare Figure\,\ref{fourheh15570}, it should be noted that we obtain very similar results with frequencies close to 1.37 and 2.37\,$d^{-1}$ (corresponding respectively to time scales of 17.5 and 10.1\,h). This is not unexpected as the corresponding peaks have very similar amplitudes. Our temporal sampling seems inadequate to discriminate between these two frequencies. When considering the complete time series, the time scale of about 17.5\,h is clearly preferred. However, it should be noted that the power spectrum of the complete time series is strongly affected by non physical peaks (see above) which could bias the relative amplitudes of other peaks. We finally note that the residual power spectra shown in Figure\,\ref{fourheh15570} present families of low amplitude peaks at frequencies slightly different from those of the main family reported above.

\begin{figure*}
\centering
\includegraphics[width=170mm]{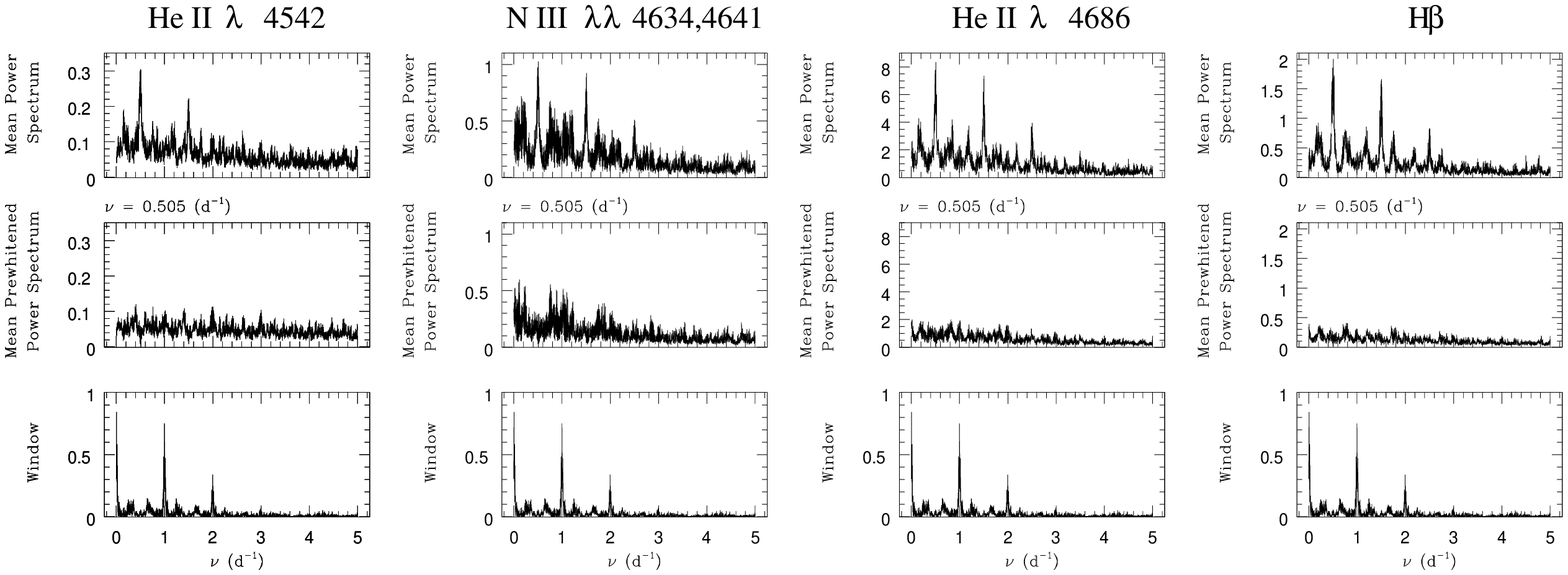}
\caption{Fourier analysis of the complete time series of HD\,16691. From the left to the right, the lines investigated are He\,{\sc ii} $\lambda$ 4542, N\,{\sc iii} $\lambda\lambda$ 4634,4641, He\,{\sc ii} $\lambda$ 4686 and H\,$\beta$. In each case, the meaning of the three panels is the same as described in Fig.\,\ref{fourhe14947}. The width of the peak at 0.505\,d$^{-1}$ is somewhat larger than that of the peaks found in the spectral window, because it is blended with that of the $1-\nu$ alias.}
\label{four16691}
\end{figure*}

Considering the rather short time scales reported above, one may wonder whether spectra collected during a same night show clear variations. The TVS technique applied to reduced times series containing 4 spectra spread over maximum 6 hours revealed indeed significant variations for most of the nights where several spectra were obtained. This lends some support to the idea that at least a part of the behaviour of the line profiles of HD\,15570 is ruled by a process occurring on time-scales shorter than a day.

\subsection{HD\,16691}\label{res3}

We detect a significant variability in the profiles of all the strongest lines in the blue spectrum, as well as in H\,$\alpha$. The TVS presents a triple peaked structure, whilst a double peaked structure was observed for the two other stars investigated in our study. The most spectacular variations are observed in the case of He\,{\sc ii} $\lambda$ 4686 and H\,$\alpha$. In the case of the He\,{\sc ii} $\lambda$ 4686 and H\,$\beta$ lines, the variability reaches its highest level in the red part of the profile. The variations in the profile of these two lines are remarkably correlated. The inspection of the spectra suggests that at least two components are present in this He\,{\sc ii} line (see individual profiles in Figure\,\ref{tvs}). However, the asymmetry of the profile remains unchanged all along our time series in the sense that the more intense part of the profile is always found on the red side. It should be noted that the variations observed in He\,{\sc ii} $\lambda$ 4686, H\,$\beta$ and H\,$\alpha$ are very similar to those observed in the case of HD\,14947 and HD\,15570, but with a much higher amplitude. The investigation of the N\,{\sc iii} lines at 4634 and 4641\,\AA\, strongly suggests that two doublets are present. In order to illustrate this, we have overplotted the N\,{\sc iii} complex profile observed on HJD\,2\,453\,295.616 (\#4 in Table\,\ref{log}) to a synthetic profile made of two shifted pairs of Gaussians superimposed on a broad bump similar to that observed in actual data. The two horizontal lines below the profile in the upper panel of Figrure\,\ref{nitrogen} illustrate the wavelength interval between two Gaussians in a given pair. This interval has been fixed at a constant value of about 7\,\AA\, in all fits, in agreement with the expected separation between N\,{\sc iii} lines respectively at 4634 and 4641\,\AA\,. The agreement between the observed profile and the synthetic one strongly suggests that two shifted N\,{\sc iii} doublets are present, or maybe that the two N\,{\sc iii} lines present a double peaked shape similar to that expected e.g. for a rotating disk.

The power spectrum of the complete time series (64 blue spectra) reveals a dominant peak at a frequency of 0.505\,d$^{-1}$ (corresponding to a time scale of 1.98\,d) for He\,{\sc ii} $\lambda$ 4542, N\,{\sc iii} $\lambda$ 4634,  N\,{\sc iii} $\lambda$ 4641, He\,{\sc ii} $\lambda$ 4686 and H\,$\beta$ (see Figure\,\ref{four16691}). We repeated the analysis on the 2007 data set (46 blue spectra) and we obtained periodograms dominated by the same frequency. Prewhitening with this frequency yields excellent results, as can be seen in Figure\,\ref{four16691} for the complete data set, except for the N\,{\sc iii} lines where significant residuals are present in the middle panel\footnote{The residual power is however much lower when the 2007 data set is considered. Because of the large emission bump present between 4600 and 4720\,\AA\,, this wavelength interval is quite sensitive to normalization errors. Slight deviations in the level of this emission bump may have an impact on the result of the Fourier analysis by introducing some additional -- and not physical -- signal in the power spectrum.}. We note that frequencies close to this one were already detected, though with some ambiguity, using the data set made of spectra from 2004 to 2006 (only 18 spectra). The gathering of more data, mostly with a better sampling of short time-scales, tends to confirm that a stable clock with a period of about 2\,d is ruling the strong variability observed in the blue spectrum of HD\,16691.

\section{Discussion} \label{disc}

\subsection{Description of the variability}\label{desc}

\subsubsection{Preliminary summary}\label{prelsum}

Considering the results presented in Section\,\ref{res}, we can formulate a series of remarks about the behaviour of our three sample stars. The various points worth noting can be separated in similarities and differences. First, we consider the following common features in the temporal behaviour of the three stars, in addition to the similarities of the mean spectrum already enumerated in Section\,\ref{spec}:
\begin{enumerate}
\item[-] the main variations are observed for He\,{\sc ii} $\lambda$ 4686, H\,$\beta$ and H\,$\alpha$.
\item[-] the variations of He\,{\sc ii} $\lambda$ 4686 consist mainly in an alternating amplification and reduction of its asymmetry.
\item[-] a similar variation of the profile asymmetry is observed in the absorption component of H\,$\beta$, with a multi-component TVS whose amplitude is stronger on the red side.
\item[-] the variations in the profile of He\,{\sc ii} $\lambda$ 4686 and H\,$\beta$ are correlated, both in time-scale and in morphology (the amplification of the asymmetry occurs at the same time in both lines).
\end{enumerate}

On the other hand, significant differences are found in the behaviour of the spectral lines investigated in this study:
\begin{enumerate}
\item[-] HD\,16691 is clearly the most variable star of our sample, both in the amplitude of the variations and in the ubiquity of its variability. Significant variations are indeed observed in all the prominent lines we investigated in the blue spectrum, and in H\,$\alpha$. Its TVS presents a central contribution in addition to the two other ones found also in the other stars of our sample.
\item[-] HD\,14947 presents also some significant variations in the N\,{\sc iii} emission lines, but the same lines in the spectrum of HD\,15570 seem to be quite stable.
\item[-] the amplitude of the variations in the asymmetry of the He\,{\sc ii} $\lambda$ 4686 and H\,$\beta$ lines increases from HD\,14947 to HD\,15570, to HD\,16691. The latter one, in its extreme configuration, suggests a partial deblending of two main emission components contributing to the profile.
\item[-] the profile of the N\,{\sc iii} emission lines is rather confused in the case of HD\,16691, and can be modelled as consisting of the blending of two pairs of lines.
\item[-] in the case of HD\,16691, a stable and well-established variability time-scale is determined ($\sim$\,2\,d). For the other two stars, a significant part of their behaviour appears to be determined by time-scales of about 7.2\,d and 17.5\,h (or perhaps 10.1\,h) respectively for HD\,14947 and HD\,15570. However, these time-scales do not seem to represent completely the behaviour of their line profiles.
\end{enumerate}

\subsubsection{Multi-Gaussian fittings}\label{disc1}
When confronted to a rather complex line profile variability, it may be a good starting point to disentangle the various components likely to contribute to these profiles. From this, the line profile variability may be investigated by varying some of these components in position or in intensity, in order to check what is the impact of this varying component on the line profile.

To perform such a simple morphological description of the line profile (without hydrodynamic or radiative transfer code), we started with the following hypotheses:
\begin{enumerate}
\item[-] the line profiles can be split into a limited number of Gaussian components,
\item[-] the parameters of each individual components can be tuned to reproduce the observed line profiles (absorption, emission, or P\,Cygni profiles).
\end{enumerate}

\begin{figure}
\centering
\includegraphics[width=80mm]{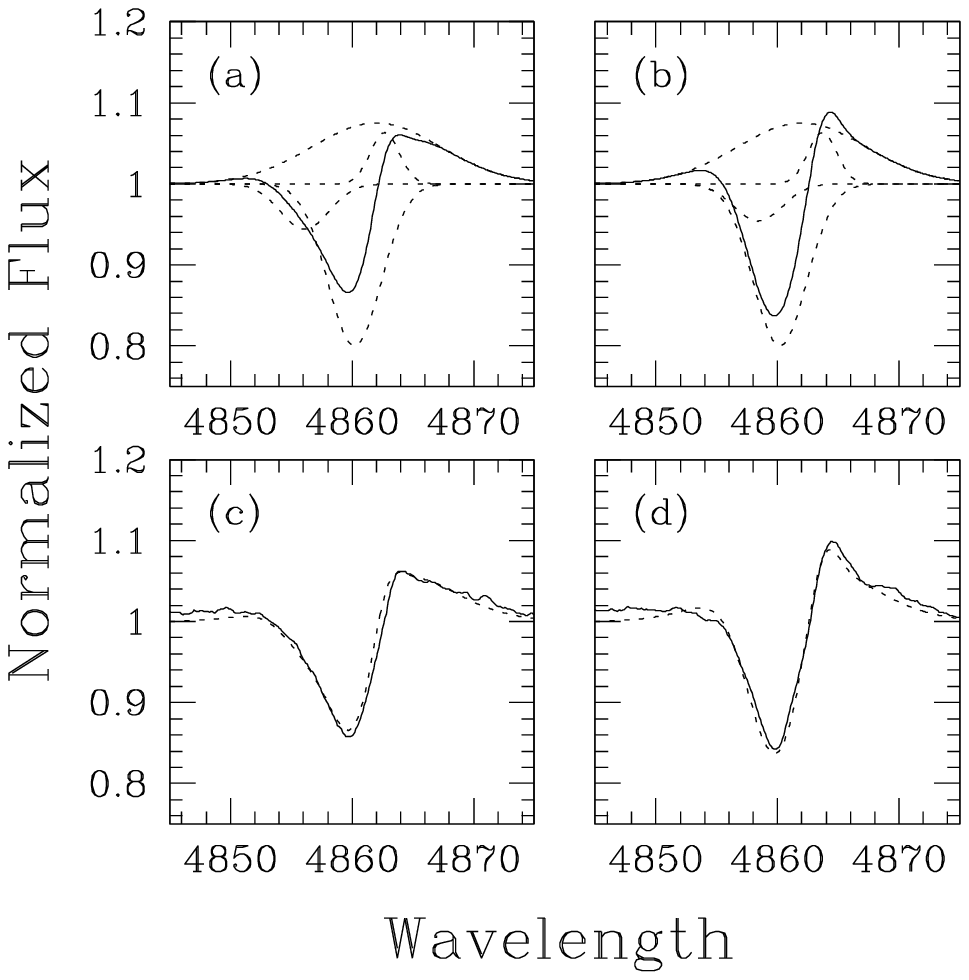}
\caption{Multi-Gaussian fitting of the H\,$\beta$ profile of HD\,16691 at two different epochs (spectra \#1 and \#32 in Table\,\ref{log}). {\it Panel (a):} Synthetic profile with its individual components plotted as dashed lines for spectrum \#1. {\it Panel (b):} same as panel (a) but for spectrum \#32. {\it Panel (c):} spectrum \#1 with the synthetic profile overplotted. {\it Panel (d):} same as panel (c) but for spectrum \#32. The wavelengths are expressed in \AA\,.\label{fithbeta}}
\end{figure}

\begin{figure}
\centering
\includegraphics[width=80mm]{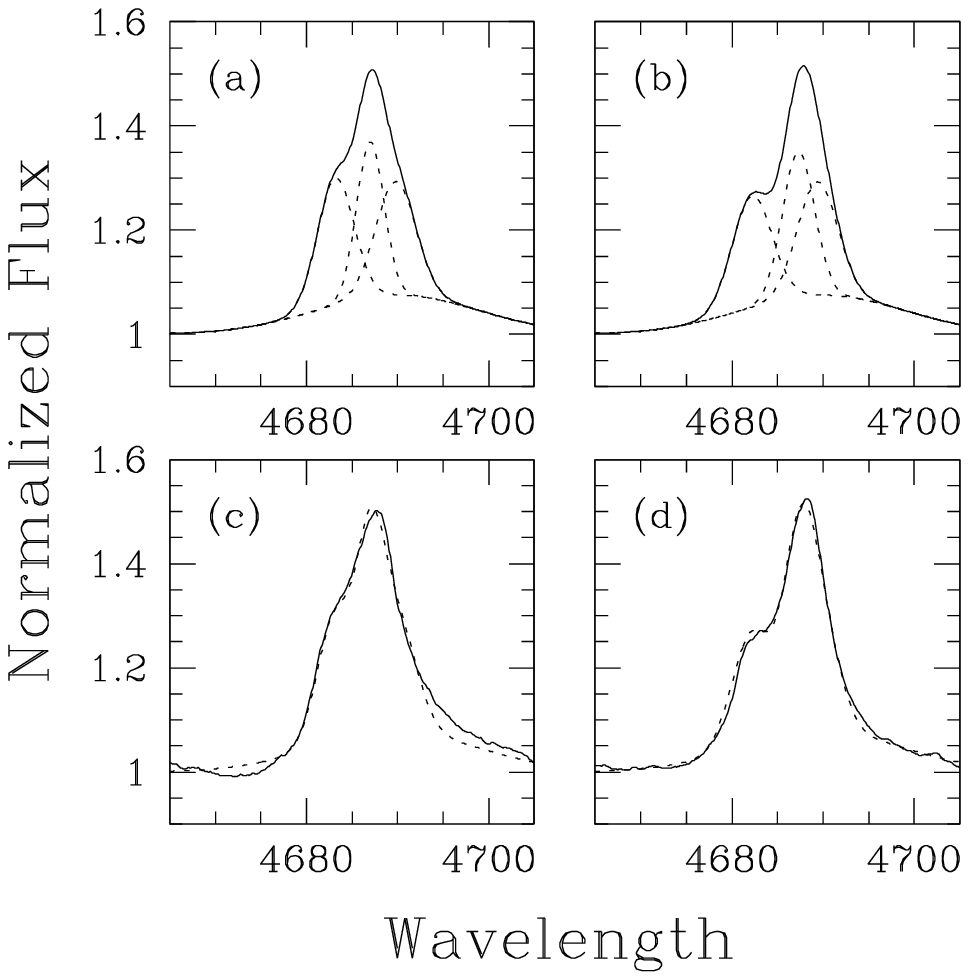}
\caption{Same as Figure\,\ref{fithbeta}, but for He\,{\sc ii} $\lambda$ 4686.\label{fitheii}}
\end{figure}

In the case of H\,$\beta$, the P\,Cygni profile is clearly dominated by its absorption component, except perhaps for HD\,16691. In order to produce synthetic profiles similar to those observed for the stars considered in this study, we used a mathematical function such as Equation\,1 in \citet{ic1805_2}, with the required number of Gaussians. For the P\,Cygni profile, we used two Gaussians shifted in wavelength with opposite signs. The H\,$\beta$ profile morphology of HD\,14947 and HD\,15570 can at first sight be reproduced using a P\,Cygni component and an additional absorption (or P\,Cygni with a weak emission component) to reproduce its asymmetry. For HD\,16691, there is a clear additional emission component seen on top of the profile in some spectra (see e.g. \#32 in the lower right panel of Figure\,\ref{tvs}). The result of the fit is shown in Figure\,\ref{fithbeta}. This set of components fits the morphology of the H\,$\beta$ line quite well. For the He\,{\sc ii} line of HD\,16691, we fitted the profiles using 4 components: 3 Gaussians of similar widths and intensities representing emission components, and a broad emission component used to account approximately for the presence of the large underlying emission bump. We note that it was necessary to adapt the normalization factors of two components (by about 10\,$\%$, sometimes more) between the two configurations to obtain acceptable fits. The results of these fits are shown in Figure\,\ref{fitheii}. In the case of HD\,14947 and HD\,15570, we reproduced quite well the profiles using the same approach, but with somewhat lower amplitude spectral shifts of the Gaussians from one spectrum to the other.

A strong contribution of the observed variations can be reproduced to some extent by varying the position of the lateral Gaussians by at most a few tens of km\,s$^{-1}$. However, the normalization of the Gaussians had also to be adapted from one spectrum to the other at a level that is sometimes significantly larger than the expected level of variations related to normalization errors in that region of the blue spectrum. We caution however that the observed profiles are only approximately represented with Gaussians. Significant deviations from this simple representation are indeed observed in many spectra.

\subsubsection{Minimum profile analysis}\label{minprof}
We built minimum emission spectra for each star by assigning to each spectral resolution element the minimum flux value of the corresponding spectral bin among the spectra of the time series. We then substracted the minimum He\,{\sc ii} $\lambda$ 4686 and H\,$\beta$ profiles from each individual observed profile in our time series. The residuals reveal the spectral location where some excess flux is detected in the line, in addition to the minimum profile. 

\begin{figure}
\centering
\includegraphics[width=80mm]{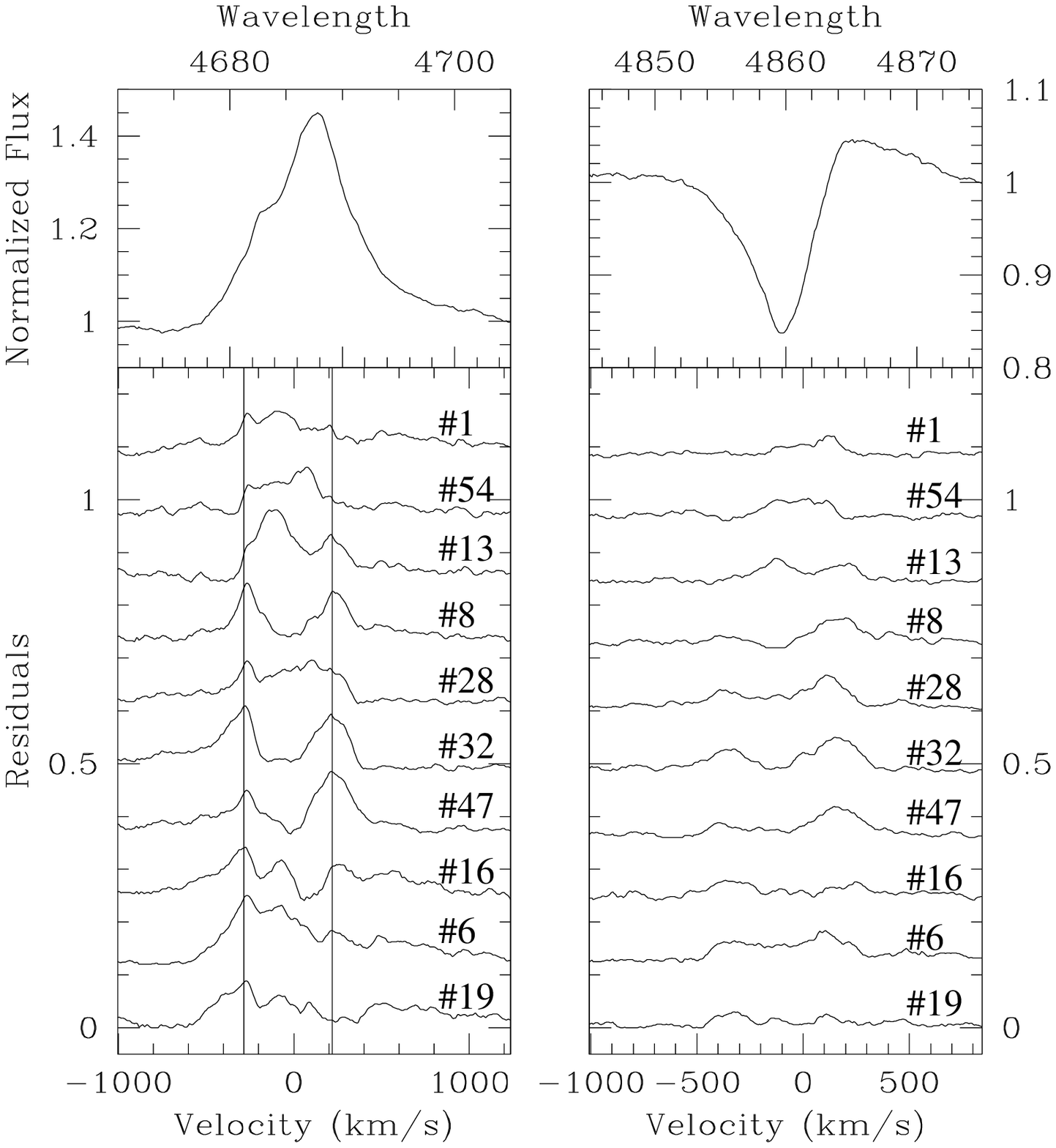}
\caption{{\it Upper panels:} minimum profiles of He\,{\sc ii} $\lambda$ 4686 {\it (left)} and H\,$\beta$ {\it (right)} of HD\,16691 as a function of wavelength expressed in \AA\,. {\it Lower panels:} residuals for a sample a 10 spectra of the time series selected in order to cover rather homogeneously the phases between 0 and 1, considering a time-scale of 1.98\,d and a T$_o$ corresponding to the time at mid-exposure of the first observation of the series. Residuals have been shifted vertically by a quantity of 0.12 units for the sake of clarity. The bottom scales corresponds to velocities expressed in km\,s$^{-1}$. The number of the spectrum as defined in Table\,\ref{log} is provided in each case. From the top to the bottom, these spectra correspond resptively to the arbitrary phases 0.00, 0.10, 0.20, 0.27, 0.45, 0.50, 0.59, 0.69, 0.81 and 0.93. In the case of He\,{\sc ii}, the two vertical lines (respectively at --280 and 220\,km\,s$^{-1}$) emphasize the alignment of the peaks observed in the residuals of most of the profiles.\label{minspec}}
\end{figure}

The most striking results are obtained for HD\,16691, as illustrated in Figure\,\ref{minspec}. A large fraction of the He\,{\sc ii} profiles display double-peaked residuals, suggesting that a variable emitting volume contributes significantly both to the blue and to the red wings of the line, on top of a minimum broad emission. The typical separation between peaks in the residuals is of the order of 500\,km\,s$^{-1}$, and their position seems to be rather stable in wavelength, with variations generally of at most 10-20\,km\,s$^{-1}$ from one observation to the other (see the vertical lines on the left part of Figure\,\ref{minspec}). However, in the case of some spectra, other excess components with higher amplitudes are also observed (see e.g. the residual \#13 in the negative radial velocity part). It is also interesting to note that at several epochs, only the blue part of the line presents a significant excess with respect to the minimum profile. We do not observe such peaks with a striking velocity confinement in the case of H\,$\beta$.

\begin{figure}
\centering
\includegraphics[width=80mm]{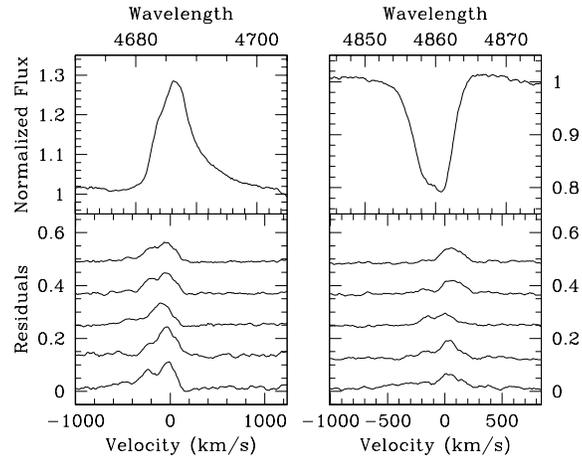}
\caption{Same as Figure\,\ref{minspec}, but for HD\,14947. The five profiles whose residuals are plotted correspond to spectra number \#1 to \#5, from the top to the bottom.\label{minspec2}}
\end{figure}

\begin{figure}
\centering
\includegraphics[width=80mm]{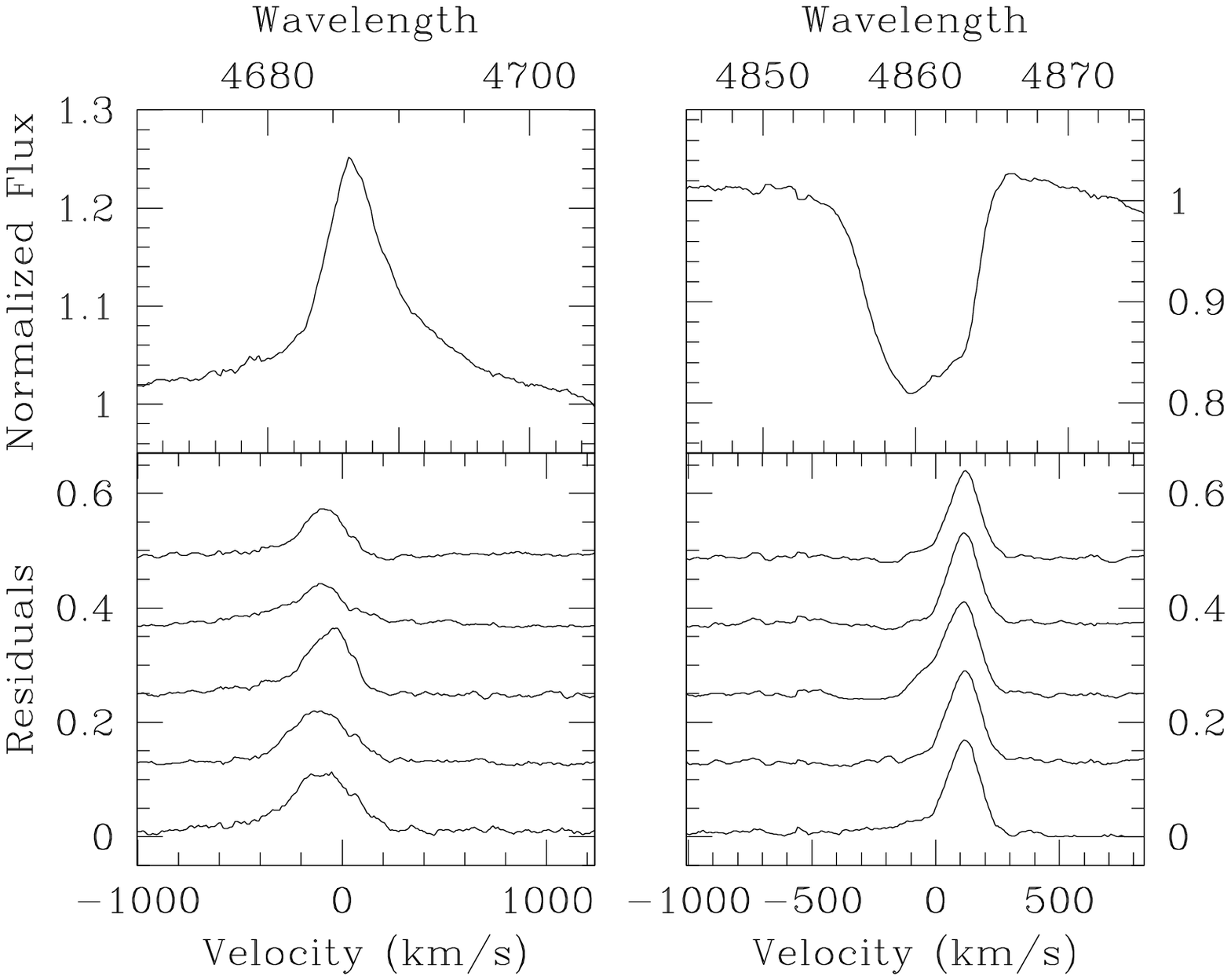}
\caption{Same as Figure\,\ref{minspec2}, but for HD\,15570.\label{minspec3}}
\end{figure}

HD\,14947 and HD\,15570 have in common to present the bulk of their excess with respect to the minimum emission profiles generally only on one side of the profile: at null or slightly negative velocities for the He\,{\sc ii} line, and at null or slightly positive velocities for H\,$\beta$ in those examaples illustrated in Figures\,\ref{minspec2} and \ref{minspec3}. We note that lower amplitudes excesses are also measured at other radial velocities, but with a much lower amplitude. This agrees with the shape of the TVS plotted in Figure\,\ref{tvs} for these two stars. This could indicate that the mmaterial that produces an extra (blueshifted) He\,{\sc ii} $\lambda$ 4686 emission simultaneously produces an extra H\,$\beta$ absorption. The main difference between these two stars comes from the fact that the residuals are more peaked and regular in the case of HD\,15570. For the latter star, the peak in the residuals of H\,$\beta$ seems to be quite stable in wavelength.

\subsection{Interpretation}
\subsubsection{A binary scenario?}
Significant line profile variations may be the signature of a binary system. For instance, we mention the case of the very massive binary WR\,20a \citep{wr20a2} whose He\,{\sc ii} $\lambda$ 4686 profiles are similar to those observed in the case of the three stars investigated in this study. In order to check the viability of such a scenario, we had a look at absorption lines less likely to be affected by wind emission to search for radial velocity excursions attributable to an orbital motion\footnote{We note that the rest wavelengths used to determine the radial velocities were taken from \citet{CLL}.}. In addition, we also searched for a periodic variation of the width of the line profile, simultaneously with a variation of its depth. As we are dealing with early-type stars, He\,{\sc i} lines are rather weak. We will thus focus on the He\,{\sc ii} line at 4542\,\AA\,.

We measured the radial velocities of He\,{\sc ii} $\lambda$ 4542 by fitting Gaussians to the line profiles. In the case of HD\,14947, we obtain a mean radial velocity for the complete time series of --32.1\,$\pm$\,7.1\,km\,s$^{-1}$. We performed a Fourier analysis of the radial velocity series and the power spectrum presents several peaks at various frequencies. The highest peak is found at 0.099\,d$^{-1}$ (corresponding to a time-scale of about 10.1\,d), with an amplitude close to 6\,km\,s$^{-1}$. Our measurements of the full width at half maximum (FWHM) of the line profile did not reveal any significant trend attributable to a binary motion. The power spectrum computed on the basis of the measurements of the depth of the line is similar to that obtained for the line profile (see Section\,\ref{res}), with the highest peak at about 0.87\,d$^{-1}$.  

In the case of HD\,15570, the multiplicity study performed by \citet{ic1805_2} did not reveal any significant radial velocity variation. We completed the latter radial velocity analysis by adding the data collected in 2007. The mean radial velocity measured on the He\,{\sc ii} $\lambda$ 4542 line is --47.1\,$\pm$\,5.8\,km\,s$^{-1}$. The Fourier analysis performed on the radial velocities revealed several peaks with amplitudes not higher than 5\,km\,s$^{-1}$. However, the Fourier analysis performed on the line depth measurements led to a power spectrum dominated by two peaks respectively at 0.369 and 1.361\,d$^{-1}$, with very similar amplitudes (the first one being the highest). These frequencies belong obviously to the family of frequencies already pointed out in Section\,\ref{res2} for this star. We did not obtain any significant result for the FWHM measurements.

Finally, we repeated the same analysis for HD\,16691. The mean radial velocity is --49.6\,$\pm$\,9.1\,km\,s$^{-1}$. None of the three quantities measured on this line presents a large amplitude variability with the same time-scale as reported in Section\,\ref{res} for the line profiles. However, the peculiar shape of the N\,{\sc iii} emission profile deserves some attention. The fact that obvious double lines are found for this profile raises the question of the multiplicity of HD\,16691. A careful inspection of the line profiles reveals that the spectral position of the components constituting the profile undergoes significant shifts, but with a rather weak semi-amplitude similar to that reported for He\,{\sc ii} $\lambda$ 4542, e.g. of the order of 10\,km\,s$^{-1}$. We fitted every N\,{\sc iii} emission profile of the complete time series in order to disentangle the four components using the same approach as shown in the left panel of Figure\,\ref{nitrogen}. It should be noted that the line intensities do not seem to be constant from one spectrum to the other. This could be due either to intrinsic changes of the components that make up the N\,{iii} lines\footnote{It should be noted that significant changes in the normalization parameters of the components used to reproduce the He\,{\sc ii} and H\,$\beta$ profiles were also requested to fit the profiles with Gaussians from one spectrum to the other.} or to changes of the underlying broad emission bump. The derived radial velocities of the Gaussian pairs are respectively of the order of --140 and 70\,km\,s$^{-1}$. The phase coverage of the time series is rather good, provided the period is 1.98\,d (see the results of the line profile temporal analysis in Sect.\,\ref{res3}), and none of the profiles of our times series shows completely merged components representative of intermediate radial velocities. The four individual components of the N\,{\sc iii} multiplet undergo only weak velocity shifts, i.e. at most 10 \,km\,s$^{-1}$ in semi-amplitude. The behaviour described in Section\,\ref{minprof}, with the double-peak shape of the residuals on top of the minimum profile of the He\,{\sc ii}\,$\lambda$\,4686 line, is similar to that of the pair of N\,{\sc iii} doublets. This situation is difficult to reconcile with the orbital motion of at least two emission components associated to two stars in a binary system. We emphasize also that no obvious signature of the presence of a companion has been found in other lines. We therefore cannot report on any signature of binarity for HD\,16691.

\subsubsection{A single star scenario?}

A plausible origin for the variations may be the stellar wind itself. Stellar winds are unlikely spherical and homogeneous. Any large scale structure in the wind may be responsible for additional emission or absorption components in spectral lines produced predominantly in the wind (e.g. He\,{\sc ii} $\lambda$ 4686 and H\,$\alpha$), or at least partly produced in the wind (e.g. H\,$\beta$). In this context, the strong correlation in the behaviour of the lines investigated in this study may be explained by their common physical origin, i.e. the same structured stellar wind.

None of the three stars present any strong evidence for binarity. Therefore, we consider a phenomenon related to a structured stellar wind, with a variability time-scale possibly related to the stellar rotation \citep[e.g.][]{oef2}. Such a phenomenon has been investigated from the theoretical point of view, e.g. corotating interaction regions \citep[see e.g.][]{cir,lobelblommecir} or a magnetically channelled wind scenario \citep[see e.g.][]{uddoula2008}.

The processes responsible for the formation of spectral lines in stellar atmospheres (stationnary or expanding) are intimately dependent on the density of their production region. Any large scale structure (with a significant density contrast with respect to the surrounding mean wind) will therefore contribute to emission lines (additional contributions due to the enhanced local density), and will have an impact on absorption lines as well (for instance, photospheric light undergoing additional absorption when crossing higher density regions). According to the density contrast and distribution, and to their physical extension, the morphological impact of such structures on spectral lines may vary substantially.\\

Large Scale Corotating Structures (LSCS) can be responsible for a substantial line profile variability. On the one hand, any deviation from the axial symmetry of the rotating emitting region will lead to periodic variations ruled by the rotation period. Such features may come from azimuthal inhomogeneities in the mass loss, or from the interaction of the wind plasma with an inclined magnetic field (the so-called {\it oblique magnetic rotator}). In this context, it is especially worth mentioning recent advances in the modelling of rigidly rotating magnetospheres \citep{RRM,RFHD}, with emphasis on the excellent results obtained in the representation of the H$\alpha$ variations in the Bp star $\sigma$\,Ori\,E \citep{rigidsigorie}. On the other hand, smaller scale heterogeneities in the stellar wind are likely to induce additional -- not necessarily periodic -- variations superimposed on those due to the LSCS. We note that we do not expect small scale density structures distributed across the whole stellar wind (i.e. clumps) to be responsible for the variations reported in this study. Such a wind clumping would more probably yield a uniform TVS, in contradiciton with the structured one presented in Figure\,\ref{tvs}. In addition, clumping is expected to lead to line profile variations with an amplitude of a few per cent at most \citep[e.g.][]{ever2}, much lower than those described in this paper. 

It should be worth checking whether the time-scales found in Sect.\,\ref{res} are compatible with the putative rotation period of the star. To do so, we used the relations (1 to 3) used by \citet{oef2} and the parameters given in Table\,\ref{paramrotper}, in order to estimate the expected minimum (constrained by the critical rotation velocity) and maximum (constrained by the projected rotational velocity) rotation periods for the three stars. The results (P$_{min}$ and P$_{max}$) are quoted in Table\,\ref{paramrotper}. In the case of HD\,16691 whose variability time-scale seems rather well-established, the 1.98\,d time-scale is compatible with the rotation period. Actually, it is rather close to the critical period below which the star would break up. On the contrary, the time-scale of 7.2\,d reported for HD\,14947 is closer to the upper limit given in Table\,\ref{paramrotper}. The case of HD\,15570 is more problematic, in the sense that the reported variability time-scales (17.5\,h, or even 10.1\,h) are significantly shorter than the break-up rotation period. However, the observed variability time scale may be a only a fraction of the actual rotation period if the symmetry of the LSCS is not rigorously cylindrical (e.g. elongated large-scale structure), or if several structures in corotation are present (e.g. structures related to several `bright' or `dark' spots on the stellar surface).

\begin{table*}
\caption{Stellar parameters of HD\,14947, HD\,15570 and HD\,16691, and estimated minimum and maximum rotation periods. The references for the stellar parameters are: (a) \citet{martins}, (b) \citet{pennyvrot}, (c) \citet{CE}, (d) \citet{markova-lpv}, (e) \citet{lei88} and (f) \citet{HPV}. \label{paramrotper}}
\begin{center}
\begin{tabular}{l c c c}
\tableline\tableline
 & HD\,14947 & HD\,15570 & HD\,16691 \\
\tableline
\vspace*{-0.2cm}\\
Sp. Type & O5If$^+$ & O4If$^+$ & O4If$^+$ \\
M$_*$ (M$_\odot$)$^{(a)}$ & 51 & 58 & 58 \\
R$_*$ (R$_\odot$)$^{(a)}$ & 19.5 & 18.9 & 18.9 \\
L$_*$ (L$_\odot$)$^{(a)}$ & 7.4\,$\times$\,10$^5$ & 8.7\,$\times$\,10$^5$ & 8.7\,$\times$\,10$^5$ \\
V$_{rot}\,\sin\,i$ (km\,s$^{-1}$) & 133$^{(b)}$ & 130$^{(c)}$ & 150$^{(c)}$ \\
V$_\infty$ (km\,s$^{-1}$) & 2300$^{(d)}$ & 2700$^{(e)}$ & 2300$^{(d)}$ \\
${\dot \mathrm{M}}$ (M$_\odot$\,yr$^{-1}$) & 1.52\,$\times$\,10$^{-5}$$^{(d)}$ & 1.78\,$\times$\,10$^{-5}$$^{(f)}$ & 1.25\,$\times$\,10$^{-5}$$^{(d)}$ \\
\vspace*{-0.2cm}\\
\tableline
P$_{min}$ (d) & 1.8 & 1.6 & 1.6 \\
P$_{max}$ (d) & 7.4 & 7.4 & 6.4 \\
\vspace*{-0.2cm}\\
\tableline
\end{tabular}
\end{center}
\end{table*}

In a scenario where the LSCS is confined by a simple, bipolar, stellar magnetic field, the apparent extension of the structure responsible for the bulk of the emission of He\,{\sc ii}\,$\lambda$\,4686 can be converted into a raw approximation of the intensity of the required magnetic field. Considering that the structure cannot be stabilized by the magnetic field beyond the Alfv\'en radius (i.e. the radius where the kinetic and magnetic energy densities equilibrate), we obtain the following relation:
$$ \mathrm{B}_*^2 = \frac{\mu_o}{4\,\pi}\,\frac{{\dot \mathrm{M}}\,\mathrm{v}_\infty}{\mathrm{R}_*^2}\,\Big[\frac{\mathrm{r}}{\mathrm{R}_*}\Big]^4\,\Big[1 - \frac{\mathrm{R}_*}{\mathrm{r}}\Big]^\beta $$
where we have assumed a radial dependence of the magnetic field (B) of the type $\mathrm{B}=\mathrm{B}_*\,(\mathrm{R}_*/\mathrm{r})^3$. In this equation, B$_*$ is the stellar magnetic field strength at the equator, B is the magnetic field strength at a distance r from the center of the star, v$_\infty$ is the terminal velocity, ${\dot \mathrm{M}}$ is the mass loss rate, $\mu_o$ is the permeability of free space, and $\beta$ is the index of the velocity law. We used a $\beta$ equal to 1. Considering that the extension of the corotating structure of HD\,16691 is of the order of 250\,km\,s$^{-1}$ in radial velocity space (on the basis of the central position of the two lateral Gaussians in Section\,\ref{disc1}) and using the quantities given in Table\,\ref{paramrotper}, we obtain a stellar magnetic field strength of about 130\,G. If we rather consider that the Alfv\'en radius corresponds to the maximum extension of the variable part of the profile (i.e. about 500\,km\,s$^{-1}$, corresponding also to the radial velocity separation between the peaks in the residuals in Figure\,\ref{minspec}), we derive a magnetic field strength of about 250\,G. For HD\,14947 and HD\,15570, the narrower variable region of the He\,{\sc ii} profile leads to somewhat lower B values. At this level of approximation, and in the context of a magnetically confined LSCS, we consider that a stellar magnetic field with a strength of a few 100\,G is compatible with our observations. 

On the basis of the above discussion, a single star scenario may provide a valuable interpretation for the spectral behaviour described in the previous sections.

\subsubsection{A common interpretation?}

As mentioned earlier in this study, several similarities are observed in the spectrum and in the temporal behaviour of the three stars we investigated. The idea of a common scenario is therefore worth considering. The differences between the lines of HD\,16691 and those of the two other stars (except for N\,{\sc iii}) are more quantitative than qualitative (i.e. similar behaviour but different amplitudes). Provided the three stars are single, we may envisage an LSCS scenario where the quantitative differences may be explained (i) by various dimensions of the large scale structure and (ii) by different inclinations of the corotating structure with respect to the line of sight, leading to different radial velocity shifts of individual emitting regions contributing to line profiles.

In all cases, a line such as He\,{\sc ii} $\lambda$ 4686 is expected to be produced by the mean wind, with substantial additional emission from the LSCS. The latter emission may appear as a double peak, in a way similar to that of a corotating emission region (see for instance Oef stars: $\zeta$\,Pup, BD\,+60$^\circ$\,2522, $\lambda$\,Cep, HD\,192281, HD\,14442, HD\,14434). The double peaked emission plus that of the mean wind can explain the complexity of the He\,{\sc ii} $\lambda$ 4686 profile. For H\,$\beta$, the main part of the profile is a classical P\,Cygni component. However, the careful inspection of the H\,$\beta$ line profiles revealed the presence of an additional absorption component moving simultaneously with at least one component of the He\,{\sc ii} $\lambda$ 4686 profile. This additional absorption may be due to the higher opacity of the LSCS at some phases of the rotation cycle. The additional small emission observed in the case of HD\,16691 may also come from the LSCS. To some extent, the slight asymmetries reported in the case of He\,{\sc ii} $\lambda$ 4542 may also be explained by the additional absorption due to the LSCS. This common origin is able to explain the strong correlation between the asymmetries of the He\,{\sc ii} $\lambda$ 4686 and H\,$\beta$ lines, as clearly shown in Figure\,\ref{tvs}. The increasing amplitude of these asymmetries from HD\,14947 to HD\,16691 may be a result of the combined effect of different inclinations with respect to the line of sight and of the spatial extension of the LSCS, resulting in the different extensions in the radial velocity space. The two N\,{\sc iii} doublets observed in the case of HD\,16691 suggest that these lines are mostly produced in the LSCS, in a confined emission region. This would imply that the region where the temperature conditions are more favorable for the production of the  N\,{\sc iii} $\lambda\lambda$ 4634,4641 is coincident with the location of the LSCS. The fact that such a feature is not observed for HD\,14947 and HD\,15570 may at least partly be explained by a lack of coincidence of the production region of the N\,{\sc iii} lines and of the LSCS. We should not reject a scenario where the LSCS is located at somewhat larger radial distances than the N\,{\sc iii} emission region in the case of HD\,14947 and HD\,15570 (explaining the single line shape and the lack of substantial variability), whereas in the case of HD\,16691 the LSCS may be extended up to much shorter distances from the star, including almost completely the N\,{\sc iii} emission region. The location and extension of the LSCS are indeed expected to depend intimately on several parameters such as wind properties and putative magnetic properties of the stars, provided the existence of the LSCS is of magnetohydrodynamic origin. In such a scenario, the variations reported on the intensity of the components contributing to line profiles (He\,{\sc ii} $\lambda$ 4686, H\,$\beta$ or N\,{\sc iii} $\lambda\lambda$ 4634,4641) may be the signature of small asymmetries in the density distribution of the LSCS. The slight variations in position measured mainly for the N\,{\sc iii} individual components, and for the residual emission in exces to the minimum profile of He\,{\sc ii}\,$\lambda$\,4686, point also to slight deviations from any cylindrical symmetry for the density distribution in the stellar wind. A large scale structure in the stellar wind modulated by the rotation of the star is therefore likely to explain most of the features observed for the three stars. 

\section{Conclusions\label{conc}} 
We presented the very first detailed investigation of the line profile variability of a sample of Of$^+$ supergiants, using unprecedented spectral time series mostly in the blue domain. We report on significant correlated variations in the profile of He\,{\sc ii} $\lambda$ 4686 and H\,$\beta$.

The stronger variations are found in the spectrum of HD\,16691 which presents some peculiarities with respect to HD\,14947 and HD\,15570. The most striking feature is the N\,{\sc iii} doublet at 4634,4641\,\AA\,, where each line seems to be double with a separation of the order of 200\,km\,s$^{-1}$. The latter feature is unlikely to be related to a binary scenario. Our temporal analysis points to a period of about 2\,d for HD\,16691, but the variability time-scale is not so well established for HD\,14947 and HD\,15570.

When put together, most of the features of the spectral behaviour of the three stars investigated in this study are compatible with a common scenario, with differences that are more quantitative than qualitative. No convincing evidence for the presence of a companion exists for the three targets, rejecting therefore variations related to binarity and colliding-wind phenomena. The behaviour described in the present study is much more compatible with a single star scenario where Large Scale Corotating Stuctures (LSCS) in the wind contribute significantly to emission and absorption features. Such a scenario appears to be well-suited for the line profile variations of He\,{\sc ii} $\lambda$ 4686 and H\,$\beta$ (and also H\,$\alpha$, even though it has not been investigated in detail in this study). In the context of this scenario, the peculiar shape of the N\,{\sc iii} $\lambda\lambda$ 4634,4641 profile in HD\,16691 is tentatively explained by a coincidence of the region that is most favorable to the production of the N\,{\sc iii} lines with a particularly extended (in radial velocity space) LSCS. The absence of such a peculiarity for the two other stars in then explained by the lack of such a coincidence.

The single star nature of the three stars investigated in this study, combined with the structured corotating wind scenario, should not be considered as a firm conclusion based on strong observational evidence, but should be rather viewed as the best working hypothesis to date for future investigations of line profile varitions of early-type stars, and in particular of Of$^+$ supergiants. 

\begin{acknowledgements}
MD wishes to express his gratitude to Dr. Eric Gosset for stimulating discussions. The travels to OHP were supported by the Minist\`ere de l'Enseignement Sup\'erieur et de la Recherche de la Communaut\'e Fran\c{c}aise. This research is also supported in part through the PRODEX XMM/Integral contract, and more recently by a contract by the Communaut\'e Fran\c{c}aise de Belgique (Actions de Recherche Concert\'ees) -- Acad\'emie Wallonie-Europe. We would like to thank the staff of the Observatoire de Haute Provence (France) for the technical support during the various observing runs. The SIMBAD database has been consulted for the bibliography.
\end{acknowledgements}


\end{document}